\documentclass[reprint,twocolumn,showpacs,showkeys]{revtex4-1}

\usepackage{latexsym,graphicx,graphics}
\usepackage{appendix}
\usepackage{amsmath,amssymb,epsfig}
\usepackage{float}

\newcommand{\abs}[1]{\left| #1 \right|}

\newcommand{\bb}[1]{ \mbox{\boldmath$ #1$}}

\newcommand{\Eq}[1]{Eq.~(\ref{#1})}
\newcommand{\Eqs}[2]{Eqs.~(\ref{#1})--(\ref{#2})}

\newcounter{enumitmpc}

\newcommand{\unit}[1]{\bb{\hat{#1}}}

\def\cFrac#1#2{%
\begin{array}{@{}c@{}}\multicolumn{1}{c|}{#1}\\%
\hline\multicolumn{1}{|c}{#2}\end{array}}

\begin{document}

\title{Waves in almost-periodic particle chains}
\author{Y. Mazor}
\surname{Mazor}
\author{Ben Z. Steinberg}
\surname{Steinberg}
\email{steinber@eng.tau.ac.il}
\thanks{ - Corresponding author. \\This research was supported by the Israel Science Foundation (grant 1503/10)}.
\affiliation{School of Electrical Engineering, Tel Aviv University, Ramat-Aviv, Tel-Aviv 69978  Israel}

\begin{abstract}
Almost periodic particle chains exhibit peculiar propagation properties that are not observed in perfectly periodic ones. Furthermore, since they inherently support non-negligible long-range interactions and radiation through the surrounding free-space, nearest-neighbor approximations cannot be invoked. Hence the governing operator is fundamentally different than that used in traditional analysis of almost periodic structures, e.g. Harper's model and Almost-Mathieu difference equations. We present a mathematical framework for the analysis of almost periodic particle chains, and study their electrodynamic properties. We show that they support guided modes that exhibit a complex interaction mechanism with the light-cone. These modes possess a two-dimensional fractal-like structure in the frequency-wavenumber space, such that a modal phase-velocity cannot be uniquely defined. However, a well defined \emph{group velocity} is revealed due to the fractal's inner-structure.
\end{abstract}



\maketitle

\section{Introduction}
Linear periodic chains of plasmonic nano-particles were studied in a number of publications \cite{Quinten}-\cite{AluPaper}. The interest stems from both theoretical and practical points of view. Particle chains were proposed as guiding structures and junctions \cite{Quinten}-\cite{EnghetaChain}, as surface waves couplers \cite{Lomakin}, as polarization-sensitive waveguides \cite{Lomakin2}, and as non-reciprocal one-way waveguides and isolators \cite{HadadSteinbergPRL}-\cite{HadadMazorSteinbergPRB}. The modal features of these periodic chains are well known, and Green's function theories revealing all the various wave constituents that can be excited in these structures were developed \cite{HadadSteinbergPRB}-\cite{HadadMazorSteinbergPRB}. Scattering due to structural disorder and its effect on the chain modes were studied \cite{AluPaper}.

Almost periodic one-dimensional (1D) structures were also studied, mainly in the context of electron dynamics in periodic magnetized crystals, or in almost periodic crystals \cite{Hofstadter,AvronSimon,BellissardSimon,SimonReview,Joaquim,AubryPaper,Jitomirsk,AvilaJitomirsk}. In these works the system dynamics is dominated by short-range interactions that naturally lead to nearest-neighbor approximations and tight-binding formulation. The resulting discrete Hamiltonian is of the general form $H\psi_n=\psi_{n+1}+\psi_{n-1}+\lambda\cos(\alpha n)\psi_n$ with irrational $\alpha/\pi$, termed as the \emph{Harper's model} (the names almost Mathieu, or almost periodic Hamiltonian are also used.) This operator is known to possess fractal (Cantor-set) spectrum. The dependence of the latter and the associated eigenfunctions, or modes, on the parameters $\alpha,\lambda$ were studied extensively.  The existence of the critical value of the modulation contrast $\lambda=2$ has been observed both theoretically and experimentally. For $\lambda<2$ the corresponding eigenfunctions are \emph{extended}, i.e.~the structure supports guided propagating modes. Beyond the critical value ($\lambda>2$) the eigenfunctions become localized and no extended modes are supported.

In carefully designed settings, these previous studies may apply also to optical systems. The works in \cite{Y_Silberberg,Zilberberg,Zilberberg2} considered 1D array of closely spaced parallel optical waveguides, arranged as an almost periodic lattice. The optical mode trapped in one waveguide may couple only to its two neighboring waveguides and cannot radiate to the free space. Therefore this system exhibits optical dynamics compliant with Harper's model. Other works on two-dimensional quasi-crystals with optical band gaps, localized modes, and directive leaky waves from slab-like domains were reported, e.g.~in \cite{QCapolino,QCapolino2,QCapolino3}.

In this work we study the propagation of optical signals in \emph{almost periodic particle chains}. The chains considered here--two typical examples of which are schematized in Fig.~\ref{fig1}--possess the following general properties. The particles are equally spaced by a distance $d$, and all possess an identical resonant frequency governed for convenience by a plasmonic-Drude model. The resonant wavelength is much larger than the particles typical size. At least one physical/geometrical property of the particles constitutes an almost periodic sequence; in Fig.~\ref{fig1}a this property is related to the (spherical) particle's volume, and in Fig~\ref{fig1}b it is related to the (ellipsoidal) particle's spatial-orientation (see details below). Due to these features, our structures differ from the previously studied ones by several important physical aspects. These differences pertain, first and foremost, to long-range \emph{vs} short-range interactions. Since the free-space dyadic Green's function describing the radiation from an excited particle decays algebraically with distance, long-range interactions between remote particles cannot be neglected and Harper's model ceases to hold. Studies of periodic chains show that the long range interactions are essential to expose the (possible) interaction of the chain with the free space radiation and the ensuing light-cone \cite{EnghetaChain}. The light-cone and radiation modes are present in our structures and play an intricate role in determining the guided modes and chain's dynamics - a mechanism absent in Harper's model. Second, the internal particle resonance plays a role in the chains spectra; it eliminates the critical passage from extended modes to localized ones. Last but not least - we show that due to the fractal nature of the chain spectra, phase velocity of the chain guided modes does not exist. However, due to the fractal's \emph{inner structure}, a definite \emph{group velocity} exists.

\section{Formulation}

\begin{figure}
    \centering
        \includegraphics[width=7cm]{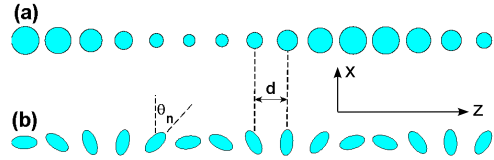}
    \caption{Examples of almost periodic particle chains. Despite the apparent long-range order, the structures never repeat themselves.}
    \label{fig1}
\end{figure}

We use the discrete dipole approximation. If an electrically small particle with electric polarizability $\bb{\alpha}$ is subject to an exciting electric field  whose local value in the \emph{absence of the particle} is $\bb{E}^L$, its response is described by the electric dipole $\bb{p}=\bb{\alpha}\bb{E}^L$. The equation governing the particle chain dynamics is
\begin{equation}
\epsilon_0\bb{\alpha}_m^{-1}\bb{p}_m=\sum_{n,\,n\ne m}\!\! {\bf A}[(m-n)d]\, \bb{p}_n.
\label{eq1}
\end{equation}
where ${\bf A}$ is the free space dyadic Green's function with source and observer restricted to the chain axis,
\begin{equation}
{\bf A}(z)=\frac{e^{ik\abs{z}}}{4\pi\abs{z}}\,\left[k^2{\bf A}_1 + \left(\frac{1}{z^2}-\frac{ik}{\abs{z}}\right){\bf A}_2\right]
\label{eq2}
\end{equation}
$d$ being the inter-particle distance, ${\bf A}_1=\mbox{diag}(1,1,0),\,{\bf A}_2=\mbox{diag}(-1,-1,2)$ and $k$ is the free space wavenumber. The polarizability of a general ellipsoidal particle is provided in \cite{HadadSteinbergPRL,MazorSteinbergPRB}.
The chain's properties are determined by the polarizabilities sequence $\{\bb{\alpha}_n\}_{n=-\infty}^{\infty}$. If it is periodic, the chain is periodic as well. Here we study the case where $\{\bb{\alpha}_n\}$ is an \emph{almost periodic} (a.p.) sequence - see \cite{LevitanBook} for a definition of quasi periodic and almost periodic sequences. Examples are shown in Fig \ref{fig1}. Fig \ref{fig1}a shows a chain of spherical particles which inverse volume (and hence $\bb{\alpha}^{-1}_n$) is modulated by an a.p. sequence; e.g. $1+\delta\cos(n)$. Another example, of further interest, is shown in \ref{fig1}b. It is a chain of ellipsoidal particles, where the $n$'th particle is rotated in the $(x,z)$ plane by $\Delta\theta$ relative to the $n-1$ particle. Then $\bb{\alpha}_n=\bb{T}_n\bb{\alpha}\bb{T}_{-n}$ where $\bb{T}_n$ is a $n\Delta\theta$ rotation operator. If $\Delta\theta/\pi$ is irrational $\{\bb{\alpha}_n\}$ is rendered a.p.
We note that every a.p. sequence $\bb{F}(n)$ of scalars or matrices may be expressed uniquely by the Fourier series
\begin{equation}
\bb{F}(n)=\sum_k \bb{a}_k e^{in\Lambda_k}
\label{eq3}
\end{equation}
The set $\{\Lambda_n\}$, called the \emph{spectrum} of $\bb{F}(n)$, is at most a countable set \cite{LevitanBook}. The additive group defined by the spectrum is termed the \emph{module} of $F(n)$, and it is denoted by the set $\{\hat{\Lambda}_r\}$.

If the sequence of matrix-polarizabilities $\bb{\alpha}_n$ is a.p., so is the sequence $\bb{\alpha}^{-1}_n$ in \Eq{eq1}. Then, from \Eq{eq3} we may represent it as,
\begin{equation}
\bb{\alpha}_n^{-1}=\sum_r\bb{a}_r e^{in\Lambda_r}=\sum_{r=-\infty}^{\infty}\hat{\bb{a}}_r e^{in\hat{\Lambda}_r}
\label{eq4}
\end{equation}
Where $\bb{a}_r$ are matrices and $\Lambda_r$ are the elements of the corresponding spectrum. The second summation is due to the fact that with no loss of information one may extend the first summation by replacing the spectrum $\{\Lambda_r\}$ by its additive group $\{\hat{\Lambda}_r\}$, and add more matrix coefficients $\hat{\bb{a}}_n$ that may or may not take the value $0$.  For convenience of subsequent derivations, the summation is indeed treated as over the entire module of $\{\bb{\alpha}_n^{-1}\}$.

Borrowing from the theory of differential equations with almost periodic coefficients \cite{CameronPaper}, the solution can be expressed as
\begin{equation}\label{eq5}
\bb{p}_n=\tilde{\bb{p}}_ne^{i\beta nd}
\end{equation}
where $\tilde{\bb{p}}_n$ is by itself an almost periodic sequence which module must be contained within the module of the almost periodic coefficients $\bb{\alpha}_n^{-1}$ in the governing formulation in \Eq{eq1} (i.e. within the set $\{\hat{\Lambda}_r\}$). This lets us write the solution as
\begin{equation}
\bb{p}_n=\left[ \sum_{\ell=-\infty}^{\infty}\bb{\Gamma}_{\ell}e^{i n \hat{\Lambda}_\ell} \right]e^{i\beta nd}.
\label{eq6}
\end{equation}
We seek a solution for the spectral vector sequence $\bb{\Gamma}_{\ell}$ and for $\beta$.  Note that the physical meaning of $\beta$ is different then seen in periodic systems, in the sense that it does not exclusively control the phase accumulation from one particle to its neighbor (or in the more general sense from one unit cell to its neighbor). This phase accumulation is also governed by the dominant frequency components in the expansion given in \Eq{eq6}.
By using \Eqs{eq4}{eq6} in \Eq{eq1} we obtain
\begin{eqnarray}
& & \epsilon_0\sum_{\ell'}\sum_{r}\hat{\bb{a}}_r\bb{\Gamma}_{\ell'}
e^{im(\hat{\Lambda}_r+\hat{\Lambda}_{\ell'})}=\\
& &\sum_{\ell}\left[ \sum_{q,\, q\neq 0}\!\!{\bf A}(qd)\bb{\Gamma}_{\ell}e^{-i\hat{\Lambda}_{\ell}q-i\beta dq}\right]e^{im\hat{\Lambda}_{\ell}}\nonumber
\label{eq7}
\end{eqnarray}
where all summations above extend from $-\infty$ to $\infty$. This equation holds several important properties which will allow further simplification in particular cases. It is an equation between two a.p.~sequences, both displayed in their corresponding formal Fourier representation. The term $\hat{\Lambda}_r+\hat{\Lambda}_{\ell'}$ is included within the $\hat{\Lambda}_{\ell}$ sequence itself. The uniqueness of these expansions \cite{LevitanBook} implies that one must require equality between the coefficients of identical frequencies. This imposes the relation $\hat{\Lambda}_r+\hat{\Lambda}_{\ell'}=\hat{\Lambda}_\ell$. We denote by $\mathbb{C}_\ell$ the set of all pairs $(r,\ell')$ that satisfy the latter relation (beware: the pairs $(r,\ell')\in\mathbb{C}_\ell$ satisfy $\ell'=\ell-r$ only in the spacial case where $\hat{\Lambda}_\ell$ is linear with $\ell$). Then \Eq{eq7} reduces to
 a difference equation for the unknown spectral vectors $\bb{\Gamma}_{\ell}$,
\begin{subequations}\label{eqs8}
\begin{equation}\label{eq8a}
\epsilon_0\sum_{(r,\ell')\in\mathbb{C}_\ell} \hat{\bb{a}}_r\bb{\Gamma}_{\ell'}-{\bf D}_{\ell}\bb{\Gamma}_{\ell}=0
\end{equation}
where ${\bf D}_{\ell}$ is a diagonal matrix, given by the summation
\begin{eqnarray}\label{eq8b}
{\bf D}_{\ell} &=& \sum_{q,\, q\ne 0}{\bf A}(qd)e^{-iq(\hat{\Lambda}_{\ell}+\beta d)}\\
               &=& \mbox{diag}(D_{\ell x},D_{\ell y},D_{\ell z}),\quad D_{\ell x}=D_{\ell y}.\nonumber
\end{eqnarray}
${\bf D}_{\ell}$ can be expressed in terms of the Polylogarithm functions $Li_s$, for which efficient summation formulas exist (see \cite{EnghetaChain,PolyLogarithmBook} and Appendix in \cite{MazorSteinbergPRB}),
\begin{equation}\label{eq8c}
{\bf D}_\ell= \frac{k^3}{4\pi}\sum_{s=1}^3 u_s f_s(kd,\beta d+\hat{\Lambda}_\ell)\bb{A}_s,
\end{equation}
where $(u_1,u_2,u_3)=(1,-i,1), \bb{A}_3=\bb{A}_2$ and
\begin{equation}\label{eq8d}
f_s(x,y)=x^{-s} [Li_s(e^{ix+iy})+Li_s(e^{ix-iy})],
\end{equation}
\end{subequations}
and where $Li_s(z)\equiv\sum_{n=1}^\infty \frac{z^n}{n^s}$ is the $s$-th order Polylogarithm function.

\section{Analysis and examples}\label{SecANALYSIS}

The spectral domain formulation in \Eqs{eq8a}{eq8d} governs the chain dynamics. Generally, it is not a tight-binding formulation, and its properties depend on the sequence $\{\hat{\bb{a}}_r\}$ which, in turn, is determined by the sequence of polarizabilities $\{\bb{\alpha}^{-1}_n\}$. The polarizability of a general ellipsoidal particle whose principal axes are aligned with the reference cartezian system, is given by
\begin{subequations}\label{eq9}
\begin{equation}\label{eq9a}
\bb{\alpha}^{-1}=\bb{\alpha}_s^{-1}-\frac{ik^3}{6\pi\epsilon_0}{\bf I}
\end{equation}
where $\bb{\alpha}_s$ is the non-radiating (``static'') component of the polarizability, obtained from
\begin{equation}\label{eq9b}
\bb{\alpha}_s^{-1}=(\epsilon_0V)^{-1}\left(\bb{\chi}^{-1}+{\bf L}\right)
\end{equation}
\end{subequations}
and where $\bb{\chi}$ is the particle material susceptibility, ${\bf I}$ is the identity matrix, and $V$ is the particle volume. ${\bf L}=\mbox{diag}(N_x,N_y,N_z)$ where $N_u$ are the depolarization factors that are given by elliptic integrals. These factors depend only on the ratios between the principal axes, and satisfy $\sum_uN_u=1$ \cite{Sihvola}. It is important to emphasize that the expression in \Eq{eq9a} includes radiation loss via the last imaginary term; i.e. it takes into account the fact that the particle may radiate into the free space around it and loose energy. Note that for deep sub-wavelength particle this term is geometry independent.

Finally, we note that generally $\bb{\chi}=\bb{\chi}(\omega)$. Hence the particle possesses a resonance frequency $\omega_r$ whenever
$\Re\{\bb{\alpha}_s^{-1}\}=0$, or
\begin{equation}\label{eq10}
\Re \{\bb{\chi}^{-1}(\omega_r)\}=-{\bf L}.
\end{equation}
For simplicity, in this work we assume an isotropic Drude model for $\bb{\chi}(\omega)$.

Below we consider two specific examples of a.p.~$\{\bb{\alpha}^{-1}_n\}$ and study the corresponding chain dynamics.

\subsection{The scalar case: spherical particles} \label{SecANALYSIS_SCALAR}

Here we examine a chain of spherical particles, as shown in Fig.~\ref{fig1}(a), with the modulated volume $V^{-1}_n=V^{-1}[1+\delta \cos(n\Delta\theta)]$ where $\Delta\theta/\pi$ is irrational. Since we have here ${\bf L}=(1/3){\bf I}$, $\bb{\alpha}_n^{-1}\mapsto \alpha_n^{-1}=\alpha_{s\, n}^{-1}-ik^3/(6\pi\epsilon_0)$ become scalars.  The Fourier series for the sequence $\alpha_n^{-1}$--i.e.~the first series in \Eq{eq4}--contains only 3 non-zero coefficients so $\alpha_n^{-1}$ may be written as
\begin{subequations}\label{eqs11}
\begin{equation}\label{eq11a}
\alpha_n^{-1}=\sum_{r=-1}^{1}a_r e^{irn\Delta\theta}
\end{equation}
with coefficients
\begin{equation}\label{eq11b}
a_{-1}=a_{1}= \frac{\delta}{2}\alpha_s^{-1},\,\,
a_{0}=\alpha^{-1}
\end{equation}
where $\alpha\equiv \alpha_n|_{\delta=0}$ and $\alpha_s\equiv \alpha_{s\, n}|_{\delta=0}$ correspond to a particles whose volume modulation contrast shrinks to zero.
We note that the series in \Eqs{eq11a}{eq11b} has a finite spectrum $\{\Lambda_\ell$\} - it is the set $\{-\Delta\theta,0,\Delta\theta\}$. Hence, the module of $\alpha_n^{-1}$ is the set
\begin{equation}\label{eq11c}
\{\hat{\Lambda}_\ell\}=\{\ell\Delta\theta\}_{\ell=-\infty}^{\infty}.
\end{equation}
\end{subequations}
By using \Eqs{eq11a}{eq11c} in \Eq{eqs8} we obtain the difference equation for the $\bb{p}_n$'s spectral amplitudes $\Gamma_\ell$
\begin{equation}\label{eq12}
M\Gamma_{\ell+1}+Q_{\ell}\Gamma_{\ell}+M\Gamma_{\ell-1}=0,\quad \forall\,\ell\in\mathbb{Z}.
\end{equation}
Here $M=\epsilon_0a_{1}$ and $Q_{\ell}=\epsilon_0 a_{0}-D_{\ell}$. The $D_{\ell}$ coefficients are obtained from \Eqs{eq8b}{eq8d}, with $D_\ell=D_{\ell x}$ ($D_\ell=D_{\ell z}$) for transverse (longitudinal) excitation.

Since $\Delta\theta/\pi$ is irrational, the sequence $Q_{\ell}$ never repeats itself. However, we emphasize that despite the structural similarity between \Eq{eq12} and Harper's equation, the former is not a result of tight-binding approximation. Furthermore, while Harper's model governs the lattice response itself, \Eq{eq12} is written on the response's \emph{spectral decomposition}. The effects of long-range interactions in our lattices are encapsulated within the structure of ${\bf D}_\ell$ or $Q_\ell$. In addition, the fact that the equation involves the $\ell$'th spectral component plus its two neighbors $\pm\ell$ only, is due to the simplicity of the particle's polarizability; only three terms are involved in the spectral decomposition in \Eq{eq11a}. Generally, the number of spectral neighbors involved in this equation is strictly determined by the number of spectral terms in the expansion of the a.p.~sequence in \Eq{eq4}. Finally, we note that in the traditional Harper's model the index-dependent coefficient is real and is of a simple cosine form, while the present formulation is generally complex and with a more complicated dependence on $\ell$.

Equations of this type, with a general periodic or a.p.~complex $Q_\ell$ were studied in \cite{BuslaevFedotov}, where sufficient conditions for the existence of Bloch solutions for that equation were developed.
However, recall again that \Eq{eq12} is written for the \emph{spectral decomposition} of the chain modes; see \Eq{eq6}. Hence, a Bloch-wave solution of \Eq{eq12} implies a localized solution for the chain response, and vice-versa; a localized non-Bloch solution of \Eq{eq12} implies a Bloch-wave solution (i.e.~a propagating mode) for the chain response. Thus, borrowing from \cite{BuslaevFedotov}, we find that a necessary condition for the latter is,
\begin{equation}\label{eq13}
q_-\equiv\inf_\ell\abs{Q_\ell}\le 2\abs{M}.
\end{equation}
The result above defines the domain in the $\beta,\omega$ space in which Bloch-wave solutions of the original a.p. difference equation, \Eq{eq1}, exist. It is used in our numerical examples below.

It is interesting to examine the chain dynamics exactly at the particle resonance $\omega=\omega_r$ for lossless material (radiation losses are still kept). In this case $\alpha_{s\, n}^{-1}=0\,\forall n$, hence $\alpha_n^{-1}=-ik^3/(6\pi\epsilon_0)\,\forall n$; the a.p. character is lost. The chain behaves exactly as a perfectly periodic one, that always possesses a trapped mode with well defined real $\beta(\omega_r)$  \cite{EnghetaChain}. This fact can also be observed mathematically directly from the a.p. formulation in \Eqs{eq12}{eq13}, as shown in appendix \ref{AppA}. In the presence of material loss $\alpha_{s\, n}^{-1}\ne 0$ at resonance, hence formally the above results do not hold, but they may still provide an approximate solution for low loss material.

When $\omega\ne\omega_r$ or/and when material loss is present, the structure may still support Bloch modes, but their analysis and the associated dispersion are not as straightforward and transparent as the case of precise resonance. To obtain a condition for the existence of a non-trivial solution sequence $\{\Gamma_\ell\}_{\ell=-\infty}^\infty$ to \Eq{eq12}, we employ a procedure as in \cite{TamirOliner} (it is not limited to periodic medium!) and obtain a double continued fraction relation,
\begin{subequations}\label{eq14}
\begin{equation}\label{eq14a}
\mathcal{K}^{(+)}_\ell\mathcal{K}^{(-)}_{\ell-1}=1
\end{equation}
where
\begin{equation}\label{eq14b}
\mathcal{K}^{(\pm)}_\ell = \cFrac{1}{q_{\ell}} \, -\, \cFrac{1}{q_{\ell\pm 1}}\, -\, \cFrac{1}{q_{\ell\pm 2}}\, -\hdots
\end{equation}
\end{subequations}
and where $q_\ell=Q_\ell/M$.
Formally, the dispersion is obtained by the ``solution pairs'' - the pairs $(\omega,\beta d)$ for which this relation is satisfied. In appendix \ref{AppB} we prove the following important properties

\begin{enumerate}

\item\label{IND}
All solution pairs are $\ell$-independent.

\item\label{SHIFT}
If $(\omega_s,\beta_s d)$ is a solution pair, so is the pair $(\omega_s,\beta_s d+\Delta\theta)$.

\item\label{SHIFTG}
If $\{\Gamma_\ell\}_\ell$ and $\{\Gamma_\ell^\Delta\}_\ell$ are the coefficient sequences that correspond to $(\omega_s,\beta_s d)$ and $(\omega_s,\beta_s d+\Delta\theta)$ respectively, then they are related by a simple shift: $\Gamma_\ell^\Delta=\Gamma_{\ell+1}$.

\setcounter{enumitmpc}{\theenumi}

 \end{enumerate}

Generally, for each $\omega_s$ there could be more than a single wavenumber $\beta_s$. We number the latter as $\beta_s^{(n)}$. From property \ref{SHIFT} it follows that for any solution pair $(\omega_s,\beta_s^{(n)} d)$, there exist infinitely many additional solution pairs $(\omega_s,\bar{\beta}^{(n)}_{s\, m})$ where $m=\pm 1, \pm 2,\ldots$, and where $\bar{\beta}^{(n)}_{s\, m}=\beta^{(n)}_s d+m\Delta\theta$. Now let $[x]_{-\pi}^\pi$ be the modulo $2\pi$ of $x$, shifted into the interval $[-\pi,\pi)$
\begin{equation}\label{eq15}
[x]_{-\pi}^\pi\! =\! x-2\pi\left\lfloor x/(2\pi)\right\rceil
\end{equation}
where $\lfloor \cdot\rceil$ denotes the nearest integer ($\pm 0.5$ go up). The pairs $(\omega_s,[\bar{\beta}^{(n)}_{s\, m}]_{-\pi}^\pi),\, m=\pm 1,\pm 2,\ldots$ form an equivalent set of solution pairs, obtained uniquely from the infinite set discussed above.
 For each $\omega_s$ and $n$, denote the set of the infinitely many corresponding $[\bar{\beta}^{(n)}_{s\, m}]_{-\pi}^\pi$'s where $m$ roams on $\mathbb{Z}$ by $\mathbb{B}^{(n)}(\omega_s)$. Since $\Delta\theta/\pi$ is irrational, $\mathbb{B}^{(n)}(\omega_s)$ is dense in the interval  $[-\pi,\pi)$ and so is $\mathbb{B}(\omega_s)=\bigcup_n\mathbb{B}^{(n)}(\omega_s)$ (at least).  Due to the above, additional properties are observed

\begin{enumerate}

\setcounter{enumi}{\theenumitmpc}

\item\label{LC}
For any $\omega_s$ that admits a non-trivial solution, the sets $\mathbb{B}(\omega_s),\,\mathbb{B}^{(n)}(\omega_s)$ always contain points inside the light-cone ($\abs{\Re\{\beta\} }<\omega/c$).

\item\label{G}
For any such $\omega_s$, it is sufficient to find the solutions $\beta_s^{(n)}d$ within an arbitrary interval of length $\Delta\theta$ in $[-\pi,\pi)$. All other solutions are obtained by $m\Delta\theta$ shifts.

\item\label{Effbeta}
For any such $\omega_s$ and for a given $n$, the effective wavenumber associated with the coefficient $\Gamma_\ell$ is $[\bar{\beta}^{(n)}_{s\, \ell}]_{-\pi}^\pi$.

\end{enumerate}

Property \ref{LC} has far reaching ramifications. In open structures, spatial harmonics with wavenumber smaller than $k=\omega/c$ always couple to the free space around the structure. Hence, formally, there are infinitely many $\ell'$ for which $\abs{\Re\{[\bar{\beta}^{(n)}_{s\, \ell'}]_{-\pi}^\pi\} }<kd$ and the corresponding coefficients $\Gamma_{\ell'}$ leak energy out. However, recall that the sequence $\{\Gamma_\ell\}_{\ell=-\infty}^\infty$ is a non-trivial vector solution of \Eq{eq12} so it must retain the same ratio between the sequence elements, independently of the specific values of $\beta$. As a result, \emph{if} there is at least one non-vanishing coefficient $\Gamma_{\ell'}$ for which $\abs{\Re\{[\bar{\beta}^{(n)}_{s\, \ell'}]_{-\pi}^\pi\} }<kd$, then the entire modal solution would experience exponential decay. The rate of decay depends on the number of such $\ell'$'s, how deep inside the light-cone they reside, and their magnitude $\Gamma_{\ell`}$ relative to the spatial harmonics that reside outside of the light-cone. Since all the spatial harmonics must decay at the same rate (in order to conserve their relative magnitudes), all the wavenumbers $[\bar{\beta}^{(n)}_{s\, \ell'}]_{-\pi}^\pi$ should have the same imaginary part. Recall now that a localized solution for the chain mode implies an extended type Bloch-wave solution of \Eq{eq12} (for the $\Gamma_\ell$'s), which evidently must have infinitely many non-vanishing coefficients inside the light-cone. Hence, in contrary to Harper's model, all localized modes in our a.p.~chain must leak energy to the free space and cannot survive for a long time.

To contrast, recall that only solutions that provide a localized coefficient sequence $\{\Gamma_{\ell}\}_\ell$ generate chain Bloch waves. These solutions may possess the property that the wavenumbers of all non-vanishing $\Gamma_\ell$ would reside outside of the light cone, thus supporting trapped modes. Below, we look for these solutions numerically in the domain defined by \Eq{eq13}.

Finally, although the proper way to analytically define and study the dispersion relation is to use \Eqs{eq14a}{eq14b}, as done above, we found it very inconvenient numerically. Therefore, to get the dispersion we truncate the infinite matrix in \Eq{eq12} to a finite equation and solve it by searching numerically for pairs $(\omega,\beta d)$ for which the matrix is rank-deficient. Naturally, some clipping of the data occurs, meaning that we set some threshold and treat only $\Gamma_{\ell}$'s which surpass the threshold. This has no significant effect on the solution accuracy since dealing with the guided modes implies a localized nature of the coefficients as mentioned before. To summarize, we choose a`-priori frequencies within the boundaries in \Eq{eq13}, seek for solutions with diminishing values of $\Gamma_{\ell}$, and set the threshold in values well below the diminishing tail of the $\Gamma_\ell$ distribution. Typical values were below 10 orders of magnitude relative to the maximal $\Gamma_\ell$. As shown below, this approach can provide very accurate results when compared to an actual simulation of an excited particle chain.

\begin{figure*}[htbp]
\vspace*{-0.1in}
    \centering
    \hspace*{-0.1in}
        \includegraphics[width=14cm]{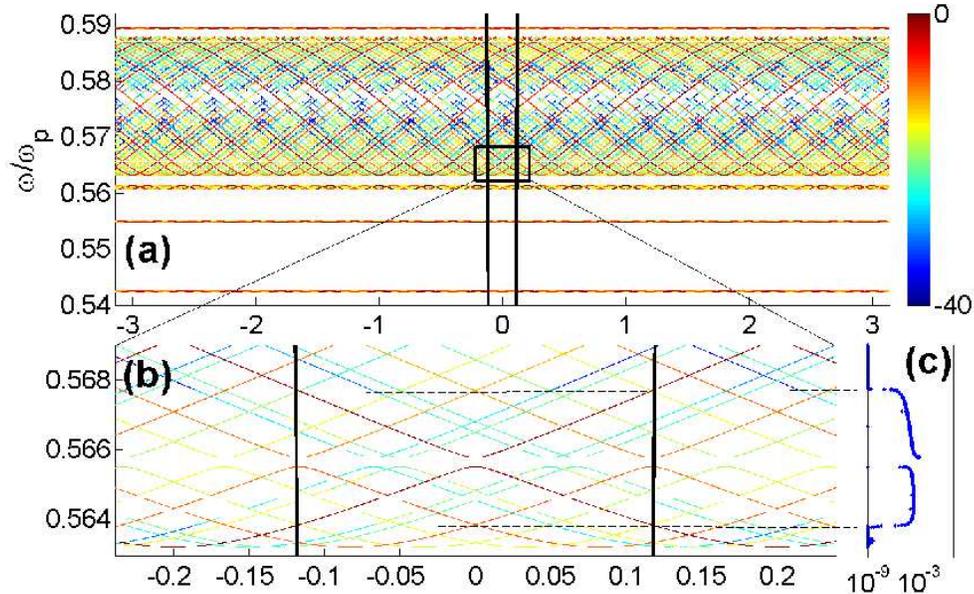}\vspace*{-0.2in}
    \caption{(a) The calculated excitation chart in the $(\omega,\Re\beta d)$ plane for $\delta=0.5,\;\Delta\theta=0.4$ Here $\ell=-45\ldots45$. Solid black lines represent the light-cone $\beta=\omega/c$. (b) Zoom-in on the frequency range selected for the numerical simulations. (c) The imaginary part of $\beta d$. Dashed lines show the frequencies for which significant transition occurs in the magnitude of $\Im\{\beta d\}$.}
    \label{fig2}
\end{figure*}

To demonstrate the properties discussed above, consider a chain with $d=\lambda_p/30$, $\delta=0.5$, and $\Delta\theta=0.4$ radians. We applied the numerical approach described above to compute the solution pairs $(\omega_s,[\bar{\beta}^{(n)}_{s\, \ell}]_{-\pi}^\pi)$--i.e.~frequencies and spatial wavenumbers--and the corresponding excitation magnitudes $\Gamma_\ell$.
The results are shown in Fig.~\ref{fig2}, color-coded according to the $\Gamma_\ell$'s magnitudes. We emphasize that although these pairs are solutions of the dispersion relation defined by \Eqs{eq14a}{eq14b}, the results should not be perceived as a ``dispersion'' in the usual sense. That is: a single point in the chart does not constitute a chain solution. Rather, all points in the charts at a given frequency are excited, each with its own excitation magnitude, in order to constitute together a wave solution. Hence, we refer to Fig.~\ref{fig2}(a) as the \emph{excitation chart}. This chart possesses a fractal-like nature in the sense that formally the band depicted in the figure is filled with solution pairs, due to the fact that the set $\mathbb{B}^{(n)}(\omega_s)$ is dense in the interval  $[-\pi,\pi)$.

The results shown in the excitation chart imply that it is impossible to define a single, or even a finite set of phase-velocities that will characterize the propagating modes along the chain. However, an inner structure of different but parallel ``branches'' is clearly observed, along which \emph{all} solution pairs are ordered. Hence, it is possible to define a group-velocity as the slope. As we show below, this uniquely defined velocity is consistent with the properties of wave-packet propagation along the chain.
The inset (b) zooms in a selected region, showing better this inner structure.

The wavenumbers $[\bar{\beta}^{(n)}_{s\, \ell}]_{-\pi}^\pi$ may be complex. Our numerical solutions for their values verified what has been predicted in the discussion following properties \ref{LC}-\ref{Effbeta}: all possess the same imaginary part.
Inset (c) shows this calculated imaginary part of $\beta d$. It is clear that when a significant branch in the excitation chart (one with color of dark-red) enters the light cone, the imaginary part increases dramatically, whereas when it resides outside the light-cone, $\Im\{\beta d\}$ is too small to be calculated precisely (we left the value $10^{-9}$ since it is the smallest value where our calculations are reasonably precise and in fact $\Im\{\beta d\}$ may become much smaller).

\begin{figure}[hp]
\vspace*{-0.1in}
    \centering
    \hspace*{-0.1in}
        \includegraphics[width=7.5cm]{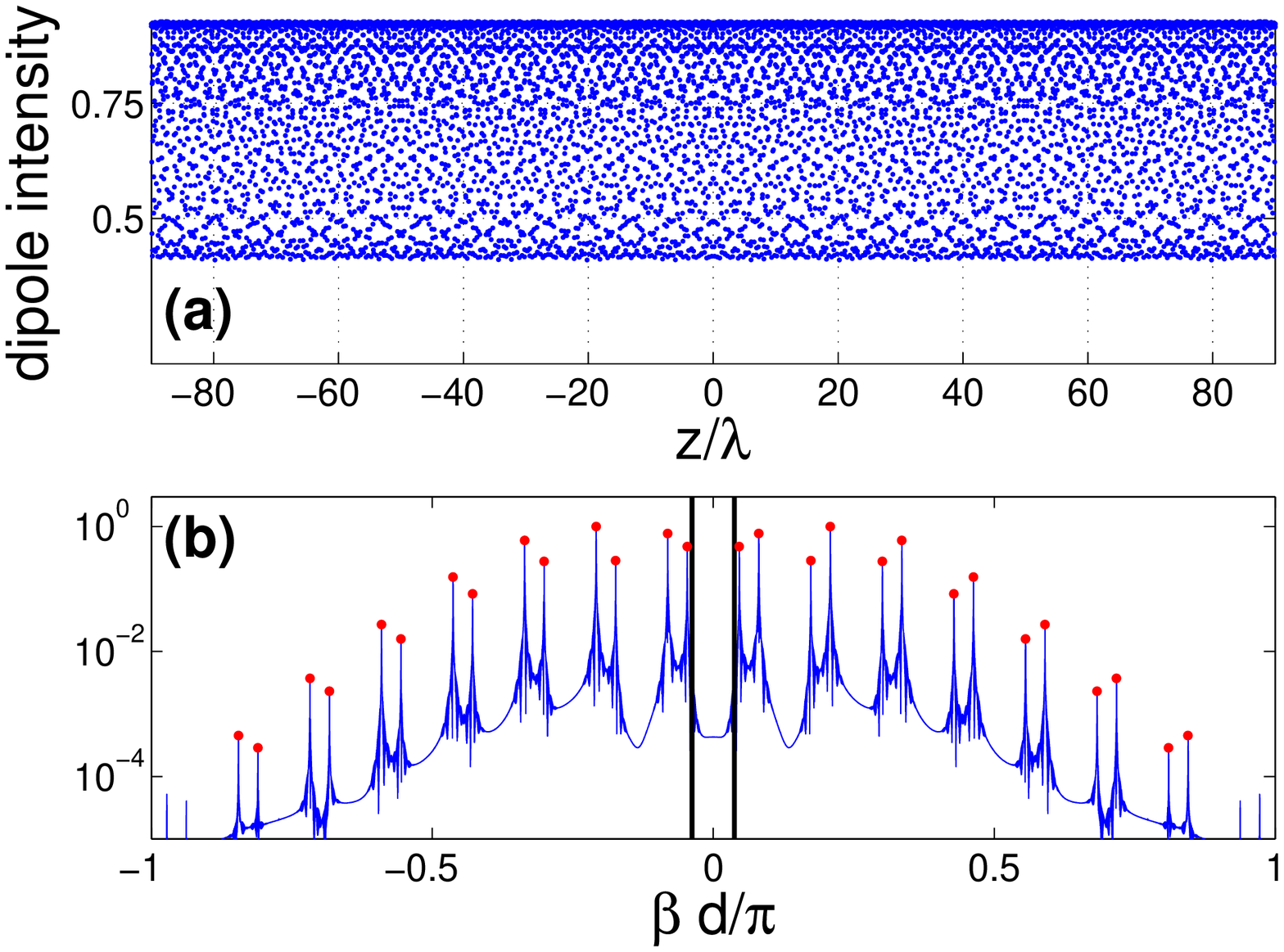}\vspace*{-0.1in}
    \caption{Response of a finite almost-periodic particle chain. (a) the dipole intensity as a function of $z$. (b) Fourier transform of the response. The red dots indicate the peaks predicted by the chart in Fig. \ref{fig2}}
    \label{fig3}
\end{figure}

\begin{figure}[hp]
\vspace*{-0.1in}
    \centering
    \hspace*{-0.1in}
        \includegraphics[width=7.5cm]{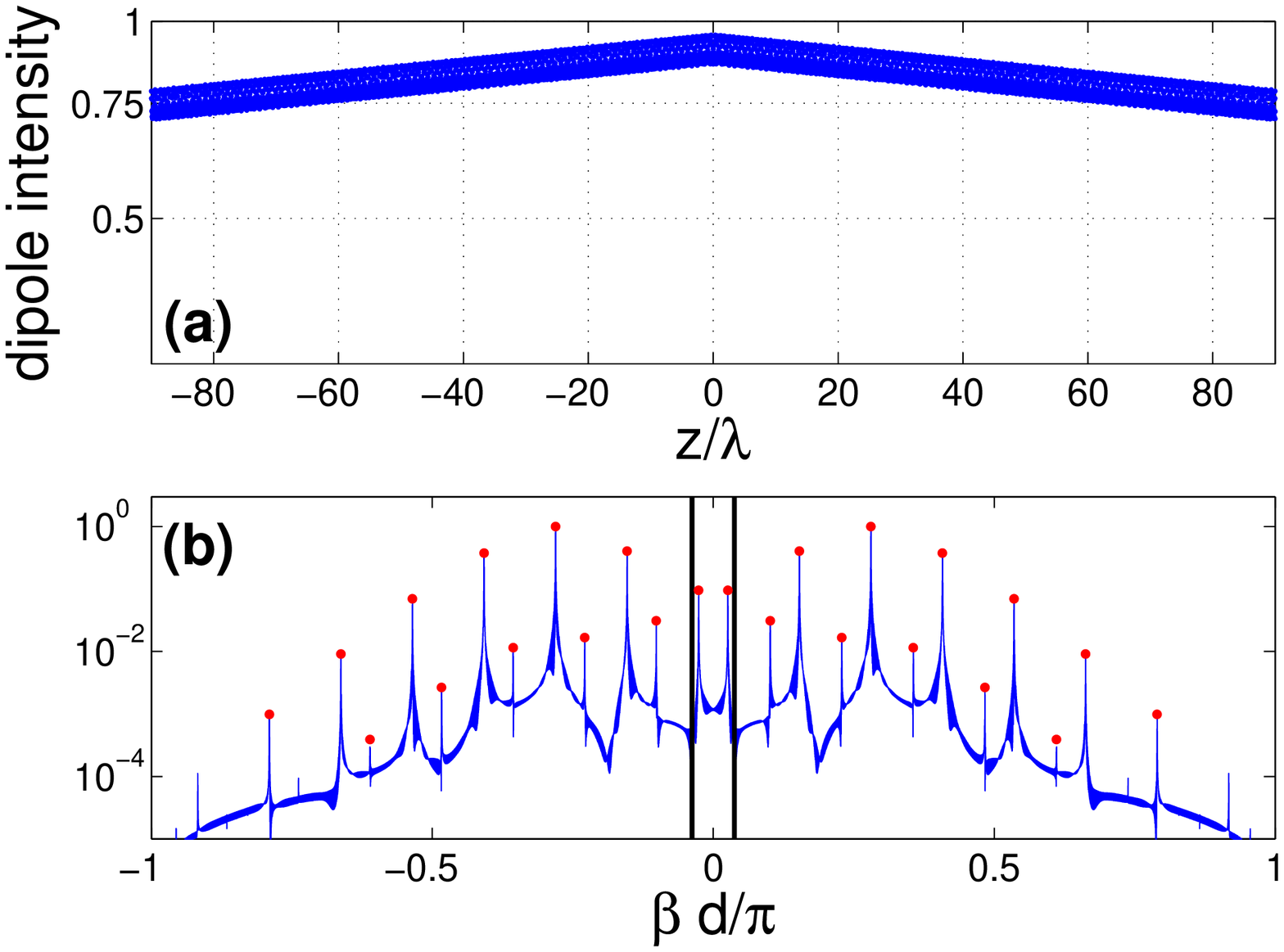}\vspace*{-0.1in}
    \caption{Response of a finite almost-periodic particle chain. (a) the dipole intensity as a function of $z$. (b) Fourier transform of the response. The red dots indicate the peaks predicted by the chart in Fig. \ref{fig2}}
    \label{fig4}
\end{figure}

To examine the validity of this chart we simulated the response of a chain of 10000 spherical particles, excited by forcing a $\unit{z}$-directed unit dipole moment on the central particle at two different frequencies (both within the inset in Fig. \ref{fig2}(b)). Figure~\ref{fig3}(a) shows the chain response at $\omega/\omega_p=0.563495$. According to the excitation chart, at this frequency all the major branches of the excitation curve are outside the light cone. Consistent with this observation, we see from Fig.~\ref{fig3}(a) that no visible attenuation is noticed. Figure~\ref{fig3}(b) shows the spatial Fourier transform of the response in Fig.~\ref{fig3}(a).
The $\Delta\theta$ spacing between the peaks is also visible. The red dots, representing the spatial harmonics as predicted by the excitation chart, along with the corresponding excitation amplitudes, also show very good agreement with the direct calculation peaks. Next, we look into the response for the frequency $\omega/\omega_p=0.567057$, displayed in Figs.~\ref{fig4}(a)-(b). According to Fig.~\ref{fig2}(b), at this frequency a significant branch of the excitation chart resides inside the light cone. Hence, although very mild, attenuation is visible in the response shown in Fig.~\ref{fig4}(a).
The value of $\Im\{\beta d\}$ predicted by Fig.~\ref{fig2}(c) is $4.09\cdot 10^{-5}$. This implies an attenuation by about $16\%$ over $80\lambda$ where $\lambda=2\pi c/\omega$. From the simulation we obtain attenuation of $15.9\%$ so the match is very good. Figure~\ref{fig4}(b) shows the Fourier transform of this response (blue), and compares it to the data predicted by the excitation chart (red dots). Again, excellent agreement is observed. Note the two peaks inside the light-cone that provide the radiation-loss mechanism that leads to the response attenuation.

\begin{figure}[htbp]
\vspace*{-0.1in}
    \centering
    \hspace*{-0.1in}
        \includegraphics[width=7.5cm]{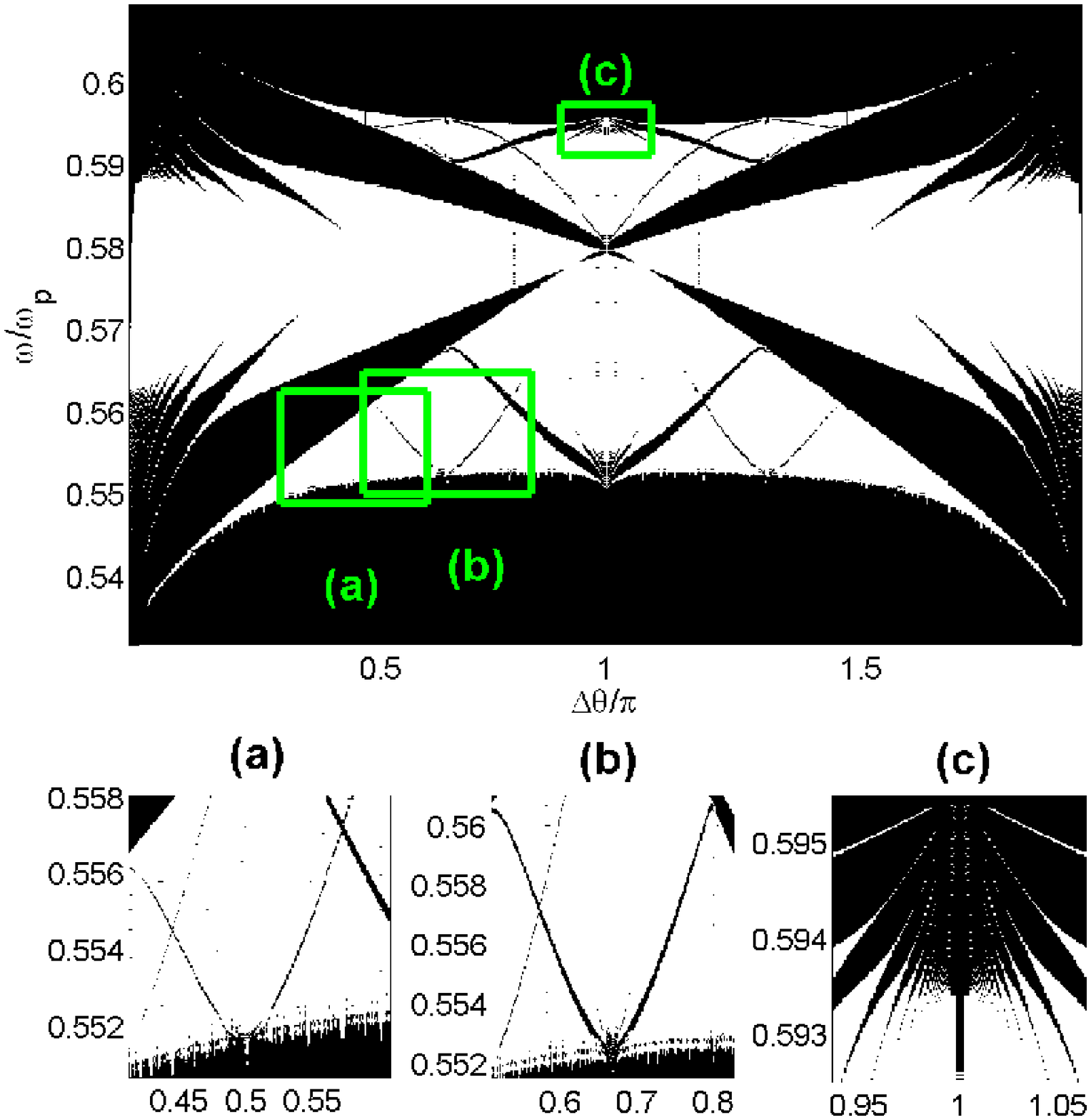}\vspace*{-0.1in}
    \caption{Excitable frequencies as a function of the modulation frequency $\Delta\theta$. White areas represent frequencies for which guided modes exist.}
    \label{fig5}
\end{figure}

Next, we examine how the modulation frequency $\Delta\theta$ affects the range of frequencies for which propagating modes may be excited.  The results are displayed in Fig.~\ref{fig5}. This plot shows a structure which has many features that resemble Hofstadter's butterfly \cite{Hofstadter}. The fractal nature is clearly visible. The frequency range for which guided modes exist, predicted by \Eq{eq13}, is $ \omega/\omega_p\in\left[ 0.5316,0.6093 \right] $ and is used as limits for the vertical axis.

\begin{figure}[htbp]
\vspace*{-0.1in}
    \centering
    \hspace*{-0.1in}
        \includegraphics[width=8.25cm]{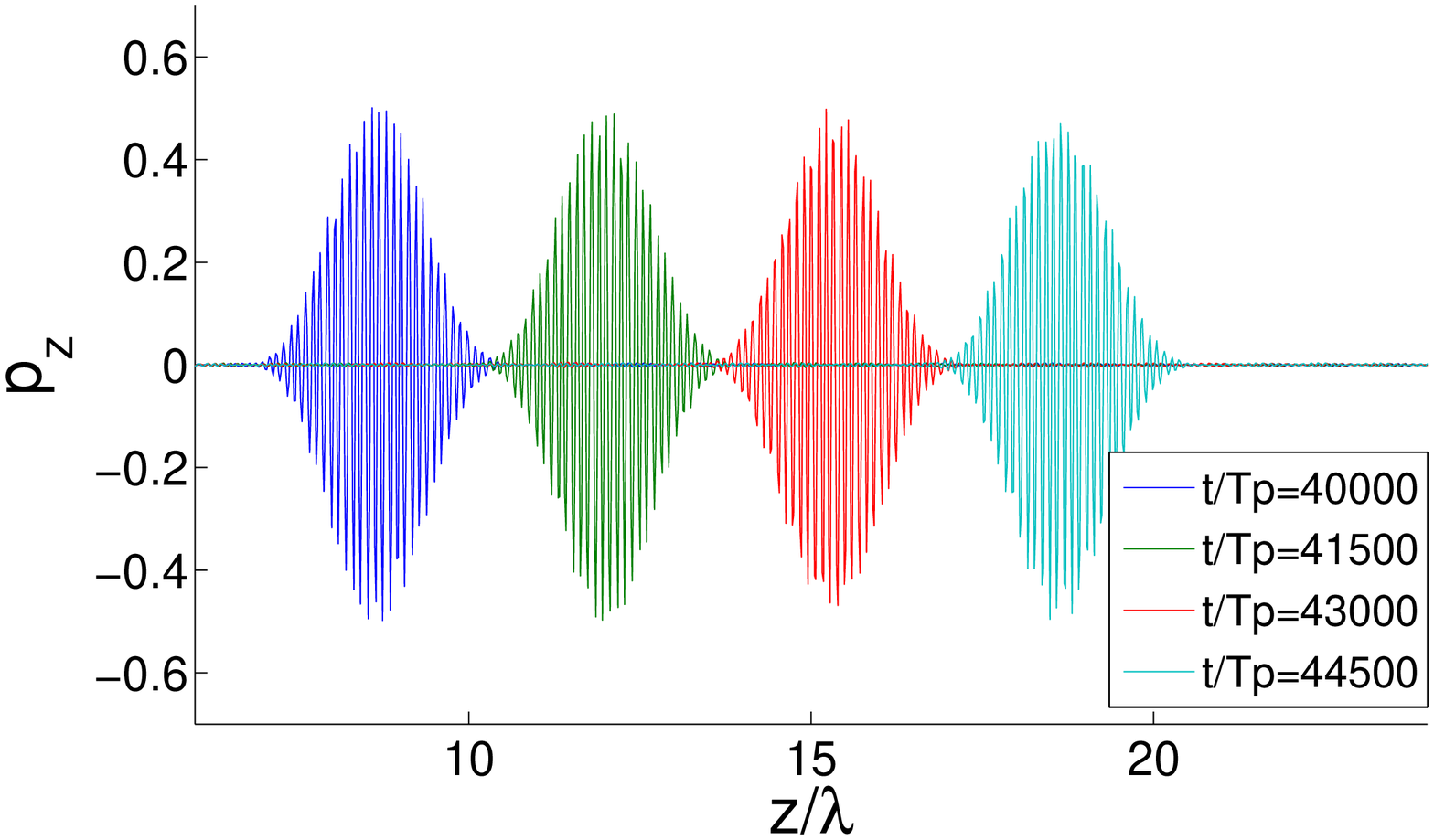}\vspace*{-0.1in}
    \caption{A pulse propagating through the a.p.~chain of modulated spherical particles.}
    \label{fig6}
\end{figure}

Finally, it is possible to excite several frequencies together, and observe the chain response to a pulse-excitation. Towards this end,
we have simulated the chain response in the time-domain due to a $\unit{z}$-directed dipole excitation of the central particle at 30 equally spaced frequencies in the range $\omega=[0.57,0.5725]\omega_p$. The frequencies are weighted by a Hamming window. This excitation creates a pulse whose temporal width is about $400 T_p$ where $T_p$ is the oscillation period of $\omega_p$. Snapshots of the chain response as a function of $z$, at four equally spaced times, are shown in Fig.~\ref{fig6}. This response shows a pulse that preserves its shape while propagating along the chain at constant velocity. This velocity is consistent with the local slopes of the inner structure revealed in Figs.~\ref{fig2}(a)-(b). Hence, as predicted, although a phase velocity cannot be defined, a uniquely defined \emph{group velocity} does exist.

\subsection{The vector case: rotating ellipsoidal particles} \label{SecANALYSIS_VECTOR}

We now turn to analyze the chain presented in Fig.~\ref{fig1}(b). This chain is a.p.~for irrational $\Delta\theta/\pi$. Many of the results reported in Sec.~\ref{SecANALYSIS_SCALAR} hold here, and particularly the formal properties \ref{IND}-\ref{Effbeta} discussed there. However, there are some important differences. First and foremost, due to the ellipsoids rotation the longitudinal and transverse polarizations are coupled, and the general matrix formulation in \Eqs{eq8a}{eq8d} cannot be reduced to a scalar one. Also, unlike the scalar case, here the ideal (lossless material) a.p.~chain doesnot possess a solution identical to that of a perfectly periodic one. Last but not least, in our numerical calculations and simulations we were not able to find a case in which a significant spatial harmonic enters the lightcone. Hence, we may conclude that the modal solutions of this chain are ``better isolated'' from the free space surrounding it, and their attenuation due to radiation loss is practically irrelevant. This observation, although based for the moment on numerical simulations, may have important practical implications.

From the almost-periodicity we again assume the solution given in \Eqs{eq5}{eq6},
which will take the full vector nature this time.  Writing $\bb{\alpha}_m^{-1}$ explicitly we obtain
\begin{equation}
\bb{\alpha}_m^{-1}=\bb{T}_{m}\bb{\alpha}^{-1}\bb{T}_{-m}
\label{eq201}
\end{equation}
where $\bb{\alpha}$ is the polarizability of the reference ellipsoidal particle and $\boldsymbol{T}_m$ is the rotation operator by $m\Delta\theta$ in the $(x,z)$ plane. The entries of this matrix are given in Appendix \ref{AlphaEllipsoids}. Note that although the angle of rotation from one particle to its neighbor is $\Delta\theta$ the spectrum of the polarizability sequence is the set $\{\Lambda_n\}=\{-2\Delta\theta,0,2\Delta\theta\}$ from which we write the module of $\bb{\alpha}^{-1}_m$ as $\{\hat{\Lambda}_r\}=\{2r\Delta\theta\}_{r=-\infty}^{\infty}$. Hence the decomposition of the sequence $\bb{\alpha}^{-1}_m$ according to \Eq{eq4} is
\begin{equation}
\bb{\alpha}_m^{-1}=\sum_{r=-\infty}^{\infty}\hat{\bb{a}}_r \, e^{ir\cdot 2\Delta\theta m}
\label{eq112}
\end{equation}
and in our specific case all the matrices $\hat{\bb{a}}_r$ are the zero matrix except $r={-1,0,1}$, for which they are identical to $\bb{a}_r$ . These matrices are also listed in Appendix \ref{AlphaEllipsoids}. Since there are again only three terms in this expansion, the dynamics formulation in \Eqs{eq8a}{eq8d} reduce to a form identical to \Eq{eq12}, but of matrix nature
\begin{equation}\label{eq18}
{\bf M}\bb{\Gamma}_{\ell+1}+{\bf Q}_{\ell}\bb{\Gamma}_{\ell}+{\bf M}\bb{\Gamma}_{\ell-1}=\bb{0},\quad \forall\,\ell\in \mathbb{Z}.
\end{equation}
where ${\bf M}=\epsilon_0\bb{a}_{1}$ and ${\bf Q}_{\ell}=\epsilon_0 \bb{a}_{0}-{\bf D}_{\ell}$.

\begin{figure*}[htbp]
\vspace*{-0.1in}
    \centering
    \hspace*{-0.1in}
        \includegraphics[width=12cm]{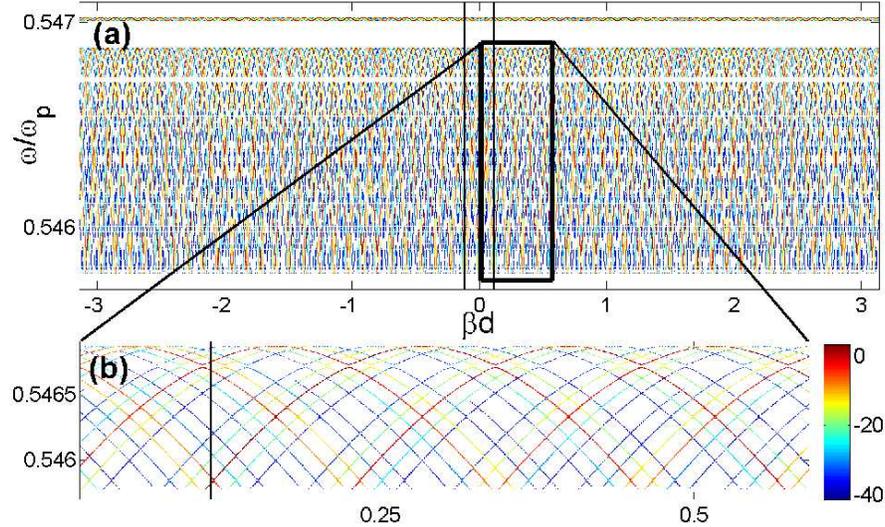}\vspace*{-0.1in}
    \caption{(a) Excitation chart for a chain of rotating ellipsoids. (b) A zoomed view.}
    \label{fig7}
\end{figure*}

For numerical example, we consider a chain with $\Delta\theta=0.4$rad, and prolate ellipsoid aspect ratio of $0.9$.
The corresponding excitation chart is shown in Fig.~\ref{fig7}. As with the scalar case, it possesses a fractal-like structure in the sense that a frequency band is filled with solution pairs; a phase velocity is hard to define. However, an inner structure of parallel lines is identified, along which all solution pairs $(\omega,\beta)$ are ordered. The corresponding slope can be associated with definite group velocity (see below). We found numerically that all the corresponding wavenumbers were real. This can be attributed to the fact that at the frequency range shown, the weight of $\abs{\bb{\Gamma}_\ell}$ that reside inside the light-cone is overwhelmed by those that reside outside it.

\begin{figure}[htbp]
\vspace*{-0.1in}
    \centering
    \hspace*{-0.1in}
        \includegraphics[width=7.5cm]{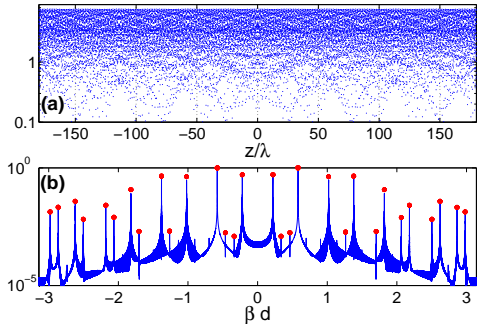}\vspace*{-0.1in}
    \caption{The resnponse of a finite chain to a dipole excitation.}
    \label{fig8}
\end{figure}

Figure \ref{fig8}(a) shows the chain response due to a $\unit{x}$-directed unit dipole excitation at $\omega=0.536445\omega_p$. No attenuation is observed over propagation distances of hundreds of wavelengths. In Fig.~\ref{fig8}(b) we show the corresponding Fourier transform (blue) compared to the data of the excitation chart (red dots). Excellent agreement is seen. Note that there are no significant peaks inside the light-cone.

\begin{figure}[htbp]
\vspace*{-0.1in}
    \centering
    \hspace*{-0.1in}
        \includegraphics[width=7.5cm]{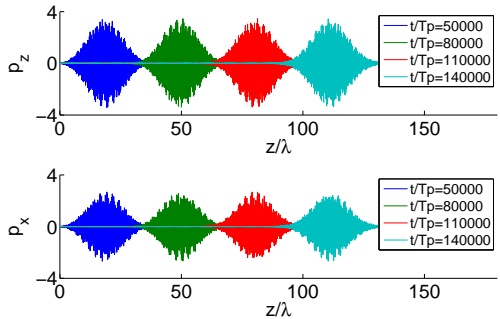}\vspace*{-0.1in}
    \caption{A pulse propagating through the ellipsoidal particle chain.}
    \label{fig9}
\end{figure}

Finally, Fig.~\ref{fig9} displays the chain response to a point dipole excitation that consists of 100 equally-spaced frequencies in the band $\omega=[0.546213,0.546324]\omega_p$, weighted by a Hamming window. A pulse that preserves its shape while propagating with a constant group velocity is observed.

\section{Conclusions}
Theoretical analysis of \emph{almost periodic} particle chains was presented, and a fractal-like dispersion relation, termed here as the \emph{excitation chart} was obtained. New chain modes existing in a.p.~particle chains were extracted and confirmed by simulations. It is shown that while phase velocity cannot be uniquely defined, these guided modes do possess a well defined \emph{group} velocity due to the inner structure of the fractal-like excitation chart. An intricate radiation mechanism that depends on the number of significant spatial harmonics inside and outside the light-cone has been observed.

\section*{ACKNOWLEDGEMENT}
This research was supported by the Israel Science Foundation (grant 1503/10).

\appendix

\section{A solution to \Eq{eq12} at resonance for lossless material}\label{AppA}

We examine \Eqs{eq12}{eq13} exactly at the particle resonance $\omega=\omega_r$ for lossless material but with radiation loss. In this case $\alpha_s^{-1}=0$ hence $M=0$, and \Eq{eq12} reduces to the requirement $Q_\ell\Gamma_\ell=0\,\forall\,\ell$. Obviously, this can be satisfied for every \emph{specific} choice of $\ell'$ provided that
\begin{subequations}
\begin{eqnarray}\label{eqA1}
\Gamma_\ell\, &=&\,\, 0\,\,\,\forall\,\ell\ne\ell'\label{eqA1a}\\
\Gamma_{\ell'} &\ne& 0,\,\,\, Q_{\ell'}|_{\omega_r}=0\label{eqA1b}
\end{eqnarray}
and the the last equation implies
\begin{equation}\label{eqA1c}
D_{\ell'}|_{\omega_r}=-ik_r^3/(6\pi)
\end{equation}
\end{subequations}
where $k_r=\omega_r/c$. Now note that \Eq{eqA1c} is identical in form to the dispersion relation of the modes of conventional periodic particle chains \cite{EnghetaChain}, thus always possesses a solution at resonance. Let $\beta_p(\omega_r)$ be that solution. Then at resonance $\beta$ of our a.p. chain must satisfy [use \Eq{eq11c} in \Eqs{eq8b}{eq8c}]
\begin{equation}\label{eqA2}
\beta d+\ell'\Delta\theta=\beta_p(\omega_r)d.
\end{equation}
Since this solution is associated with a single non-zero coefficient $\Gamma_{\ell'}$, it constitutes a Bloch solution of the a.p. chain. Furthermore, using this fact in \Eq{eq6}, we find that at resonance the solution of the corresponding periodic chain always holds for its a.p. counterpart.

\section{Properties of the continued-fraction dispersion \Eq{eq14}}\label{AppB}

First, we note that $\mathcal{K}^{(\pm)}_\ell$ satisfy the following property,
\begin{equation}\label{eqB1}
\frac{1}{\mathcal{K}^{(\pm)}_\ell}=q_\ell-\mathcal{K}^{(\pm)}_{\ell\pm 1}
\end{equation}
Now assume that \Eq{eq14a} is satisfied by a pair $(\omega_0,\beta_0)$ and rewrite it as,
\begin{equation}\label{eqB2}
\frac{1}{\mathcal{K}^{(+)}_\ell}=\mathcal{K}^{(-)}_{\ell-1}.
\end{equation}
But with \Eq{eqB1} the dispersion relation in \Eq{eqB2} can be written as
\begin{equation}\label{eqB3}
\mathcal{K}^{(+)}_{\ell+1}=\frac{1}{\mathcal{K}^{(-)}_\ell}
\end{equation}
The last equation is identical to \Eq{eqB2} subject to the shift $\ell\mapsto\ell+1$. Hence, any solution pair $(\omega_0,\beta_0)$ is $\ell$-independent, as stated in property \ref{IND} in Sec.~\ref{SecANALYSIS_SCALAR}. Furthermore, note that from \Eq{eq8c}, from \Eq{eq11c}, and from \Eq{eq12}, the dependence of $q_\ell$ on $\ell$ and $\beta$ has the form $q_\ell(\beta)=q(\beta d+\ell\Delta\theta)$. Hence, by substituting the solution pair $(\omega_0,\beta_0)$ into \Eq{eqB3}, one reconstructs the dispersion in \Eq{eqB2} but with $\beta d + \Delta\theta$. This proves property \ref{SHIFT} in Sec.~\ref{SecANALYSIS_SCALAR}. Finally, property \ref{SHIFTG} in Sec.~\ref{SecANALYSIS_SCALAR} follows directly from the above and from the dependence of $Q_\ell$ on $\Delta\theta$.

\section{The polarizability sequence of rotating ellipsoids chain and its Fourier decomposition}\label{AlphaEllipsoids}

The entries of the $\boldsymbol{\alpha}_m^{-1}$ are given by
\begin{widetext}
\begin{subequations}
\begin{eqnarray}
\alpha^i_{m,(11)} & = & \frac{\alpha^i_{xx}+\alpha^i_{zz}}{2}+\frac{\alpha^i_{xx}-\alpha^i_{zz}}{2}\cos{2m\Delta\theta}+\frac{\alpha^i_{zx}+\alpha^i_{xz}}{2}\sin{2m\Delta\theta} \label{eqC1a}\\
\alpha^i_{m,(13)} & = & \frac{\alpha^i_{xz}-\alpha^i_{zx}}{2}+\frac{\alpha^i_{xz}+\alpha^i_{zx}}{2}\cos{2m\Delta\theta}+\frac{\alpha^i_{zz}-\alpha^i_{xx}}{2}\sin{2m\Delta\theta} \label{eqC1b}\\
\alpha^i_{m,(22)} & = & \alpha^i_{yy} \label{eqC1c}\\
\alpha^i_{m,(31)} & = & \frac{\alpha^i_{zx}-\alpha^i_{xz}}{2}+\frac{\alpha^i_{xz}+\alpha_{zx}}{2}\cos{2m\Delta\theta}+\frac{\alpha^i_{zz}-\alpha^i_{xx}}{2}\sin{2m\Delta\theta}
\label{eqC1d}\\
\alpha^i_{m,(33)} & = & \frac{\alpha^i_{xx}+\alpha^i_{zz}}{2}+\frac{\alpha^i_{zz}-\alpha^i_{xx}}{2}\cos{2m\Delta\theta}-\frac{\alpha^i_{zx}+\alpha^i_{xz}}{2}\sin{2m\Delta\theta}
\label{eqC1e}
\end{eqnarray}
\end{subequations}
\end{widetext}
where $\boldsymbol{\alpha}^i_m=\boldsymbol{\alpha}^{-1}_m$ is the inverse polarizablity of the m'th particle and $\boldsymbol{\alpha}^i=\boldsymbol{\alpha}^{-1}$ is the inverse polarizability of the reference particle. The corresponding Fourier coefficients according to \Eqs{eqC1a}{eqC1e} are
\begin{widetext}
\begin{subequations}
\begin{equation}\label{eqC2a}
\boldsymbol{a}_{0}=
\begin{pmatrix}
\frac{1}{2}\alpha^i_{xx}+\frac{1}{2}\alpha^i_{zz} & 0 & \frac{1}{2}\alpha^i_{xz}-\frac{1}{2}\alpha^i_{zx} \\
0 & \alpha^i_{yy} & 0 \\
-\frac{1}{2}\alpha^i_{xz}+\frac{1}{2}\alpha^i_{zx} & 0 & \frac{1}{2}\alpha^i_{xx}+\frac{1}{2}\alpha^i_{zz}
\end{pmatrix},\quad \boldsymbol{a}_{1}=\left( \boldsymbol{a}_{-1} \right)^{*}
\end{equation}
\begin{equation}\label{eqC2b}
\boldsymbol{a}_{-1}=
\begin{pmatrix}
\frac{1}{4}\alpha^i_{xx}-\frac{1}{4i}\alpha^i_{xz}-\frac{1}{4i}\alpha^i_{zx}-\frac{1}{4}\alpha^i_{zz} & 0 & \frac{1}{4i}\alpha^i_{xx}+\frac{1}{4}\alpha^i_{xz}+\frac{1}{4}\alpha^i_{zx}-\frac{1}{4i}\alpha^i_{zz} \\
0 & 0 & 0 \\
\frac{1}{4}\alpha^i_{xz}+\frac{1}{4i}\alpha^i_{xx}+\frac{1}{4}\alpha^i_{zx}-\frac{1}{4i}\alpha^i_{zz} & 0 & -\frac{1}{4}\alpha^i_{xx}+\frac{1}{4i}\alpha^i_{xz}+\frac{1}{4i}\alpha^i_{zx}+\frac{1}{4}\alpha^i_{zz}
\end{pmatrix}
\end{equation}
\end{subequations}
\end{widetext}


\begin{thebibliography}{1}


\bibitem{Quinten}
M.~Quinten, A.~Leitner, J.~R.~Krenn, and F.~R.~Aussenegg,
 \emph{Optics Letters}, {\bf 23}(17), 1331 (1998).

\bibitem{TretyakovViitanen}
S.~A.~Tretyakon and A.~J.~Viitanen,
\emph{Electrical Engineering}, {\bf 82}, 353-361 (2000).

\bibitem{Brongersma}
M.~L.~Brongersma, J.~L.~Hartman, and H.~A.~Atwater,
\emph{Phys.~Rev.~B}, {\bf 62}(24) R16356 (2000).

\bibitem{WeberFordPRB70}
W.~H.~Weber and G.~W.~Ford,
\emph{Phys.~Rev. B} {\bf 70}, 125429 (2004).

\bibitem{KP_PRB74}
A.~F.~Koenderink and A.~Polman,
\emph{Phys.~Rev. B} {\bf 74}, 033402 (2006).

\bibitem{YangCrozierOpEx16}
T.~Yang and K.~B.~Crozier,
\emph{Optics Express} {\bf 16}(12), 8570 (2008).

\bibitem{EnghetaChain}
A.~Alu and N.~Engheta,
\emph{Phys.~Rev. B} {\bf 74}, 205436 (2006).

\bibitem{Lomakin}
D.~V.~Orden, Y.~Fainman, and V.~Lomakin,
coupled with surfaces,''  \emph{Opt.~Lett.}, {\bf 34}(4), 422-424 (2009).

\bibitem{Lomakin2}
D.~V.~Orden, Y.~Fainman, and V.~Lomakin,
\emph{Opt.~Lett.}, {\bf 35}(15), 2579-2581 (2010).

\bibitem{Capolino_Leaky}
S.~Campione, S.~Steshenko, and F. Capolino,
\emph{Optics Express},  {\bf 19}(19), 18345-18363 (2011).

\bibitem{HadadSteinbergPRL}
Y.~Hadad and Ben Z.~Steinberg,
\emph{Phys.~Rev.~Lett.}, {\bf 105}, 233904 (2010).

\bibitem{MazorSteinbergPRB}
Y.~Mazor and Ben Z.~Steinberg, \emph{Phys.~Rev.~B} {\bf 86}, 045120 (2012).

\bibitem{HadadSteinbergPRB}
Y.~Hadad and Ben Z.~Steinberg, \emph{Phys.~Rev.~B} {\bf 84}, 125402 (2011).


\bibitem{HadadMazorSteinbergPRB}
Y.~Hadad, Y.~Mazor and Ben Z.~Steinberg,
\emph{Phys.~Rev.~B}, {\bf 87}, 035130 (2013).

\bibitem{AluPaper}
A.~Alu and N.~Engheta, \emph{New J. Phys.} {\bf 12} 013015 (2010).


\bibitem{Hofstadter} 
Douglass R.~Hofstadter,
\emph{Phys. Rev. B} {\bf 14}(6), pp. 2239-2249 (1976).

\bibitem{AvronSimon}
J.~Avron and B.~Simon,
\emph{Bulletin of the American Mathematical Society}, {\bf 6}(1), 81-85 (1982).


\bibitem{BellissardSimon}
J.~Bellissard and B.~Simon,
\emph{Journal of Functional Analysis}, {\bf 48}, 408-419 (1982).

\bibitem{SimonReview}
B.~Simon,
\emph{Advances in Applied Mathematics}, {\bf 3} 463-490 (1982).

\bibitem{Joaquim}
P. Joaquim,
\emph{Commun.~Math.~Phys.} {\bf 244} 297-309 (2004).

\bibitem{AubryPaper}
S.~Aubry and G.~Andre´,
\emph{Ann. Israel Phys. Soc.} {\bf 3},133–140 (1980).

%

\bibitem{Jitomirsk}
S.~Ya.~Jitomirskaya,
\emph{Annals of Mathematics}, {\bf 150}, 1159 (1999).

\bibitem{AvilaJitomirsk}
A. Avila and S.~Ya.~Jitomirskaya,
\emph{Annals of Mathematics}, {\bf 170}(1), 303 (2009).

\bibitem{Y_Silberberg}
Y.~Lahini, R.~Pugatch, F.~Pozzi, M.~Sorel, R.~Morandotti, N.~Davidson, and Y.~Silberberg,
\emph{Phys.~Rev.~Lett.} {\bf 103}, 013901 (2009).

\bibitem{Zilberberg} 
Y.~E.~Kraus, Y.~Lahini,~Z.~Ringel, M.~Verbin, and O.~Zilberberg,
\emph{Phys. Rev. Lett.} {\bf 109}, 106402 (2012).

\bibitem{Zilberberg2}
M.~Verbin, O.~Zilberberg, Y.~E.~Kraus, Y.~Lahini, and Y.~Silberberg,
\emph{Phys. Rev. Lett.} {\bf 110}, 076403 (2013).


\bibitem{QCapolino} 
A.~Della Villa, S.~Enoch, G.~Tayeb, V.~Pierro, V.~Galdi, F.~Capolino,
\emph{Phys.~Rev.~Lett.} {\bf 94}, 183903 (2005).


\bibitem{QCapolino2} 
A.~Della Villa, S.~Enoch, G.~Tayeb, F.~Capolino, V.~Pierro, V.~Galdi,
\emph{Optics Express} {\bf 14}(21), 10021-10027 (2006).

\bibitem{QCapolino3}
A.~Micco, V.~Galdi, F.~Capolino, A.~D.~Villa, V.~Pierro, S.~Enoch, and G.~Tayeb,
\emph{Phys. Rev. B} {\bf 79}(7), 075110 (2009).






\bibitem{LevitanBook}
B. M. Levitan and V. V. Zhikov, \emph{Almost periodic functions and differential equations}, Cambridge University Press, 1983

\bibitem{CameronPaper}
R. H. Cameron,
\emph{Duke Math. J.}, {\bf 1}(3), 356-360 (1935)

\bibitem{PolyLogarithmBook}
Leonard Lewin, \emph{Polylogarithms and Associated Functions}, Elsevier , New York, 1981.

\bibitem{Sihvola}
 A. H. Sihvola, \emph{Electromagnetic Mixing Formulas and Applications,}
Electromagnetic Waves Series (IEE, London, 1999).

\bibitem{BuslaevFedotov}
V.~S.~Buslaev and A.~A.~Fedotov,
\emph{St.~Petersburg Math.~J.}, {\bf 7}(4), 561-594 (1995).

\bibitem{TamirOliner}
T.~Tamir, H.C.~Wang and A.A.~Oliner,
\emph{IEEE Trans. Microw. Theory Tech.}, {\bf 12}(3), 323-335 (1964)


\end{thebibliography}
\end{document}